\documentclass[prd,preprint,tightenlines,floatfix,showpacs,preprintnumbers,nofootinbib,eqsecnum]{revtex4}
 \usepackage[dvips,final]{graphicx}
  \usepackage{amssymb}
   \usepackage{amsmath}
    \usepackage{amsfonts}
     \usepackage{epsfig}
      \usepackage{bm}

 \newcommand\beq{\begin{equation}}
 
 \newcommand\eeq{\end{equation}}
 \newcommand\beqn{\begin{eqnarray}}
 \newcommand\eeqn{\end{eqnarray}}

\begin{document}

\begin{flushright}
LU TP 13-06\\
February 2013
\end{flushright}

\title{On a possible compensation of the QCD vacuum
contribution\\ to the Dark Energy}

\author{Roman Pasechnik}
\affiliation{Department of Astronomy and Theoretical Physics, Lund
University, SE-223 62 Lund, Sweden}

\author{Vitaly Beylin}
\affiliation{Research Institute of Physics, Southern Federal
University, 344090 Rostov-on-Don, Russian Federation}

\author{Grigory Vereshkov}
\affiliation{Research Institute of Physics, Southern Federal
University, 344090 Rostov-on-Don, Russian Federation}
\affiliation{Institute for Nuclear Research of Russian Academy of
Sciences, 117312 Moscow, Russian Federation\vspace{1cm}}

\begin{abstract}
\vspace{0.5cm} We suggest one of the possible ways to compensate the
large negative quantum-topological QCD contribution to the vacuum
energy density of the Universe by means of a positive constant
contribution from a cosmological Yang-Mills field. An important role
of the exact particular solution for the Yang-Mills field
corresponding to the finite-time instantons is discussed. An
interesting connection of the compensation mechanism to the color
confinement in the framework of instanton models has been pointed
out. Besides the $\Lambda_{\rm QCD}$ scale, this proposal relies on
one yet free dimensionless normalisation constant which cannot be
fixed by the perturbative QCD theory, and thus should be fine-tuned
for the exact compensation to hold.
\end{abstract}

\pacs{98.80.Qc, 98.80.Jk, 98.80.Cq, 98.80.Es}

\maketitle

\section{Introduction}

Current accelerated expansion of the Universe is commonly attributed
to the existence of the so-called Dark Energy which is confirmed in
many cosmological observations so far, e.g. in studies of the type
Ia Supernovae \cite{SNE1A}, cosmic microwave background anisotropies
\cite{WMAP}, large scale structure \cite{SDSS} etc. The Standard
Cosmological Model is based on the time-independent Dark Energy
approximation called the cosmological constant, or $\Lambda$-term,
approximation which agrees well with current observational data.
However, the problem of theoretical interpretation and prediction of
fundamental properties of the Dark Energy (or the $\Lambda$-term)
remains one of the major unsolved problem of Theoretical Physics
\cite{Weinberg:1988cp}. For a comprehensive overview of existing
theoretical models and interpretations of the Dark Energy, see e.g.
Refs.~\cite{Copeland:2006wr,Bamba:2012cp,Bamba,Li:2012dt,Yoo:2012ug}
and references therein.

One of the traditional interpretations of the $\Lambda$-term is by
means of the vacuum energy density satisfying the equation of state
$P_{\Lambda}=-\Lambda$ with vacuum pressure $P_{\Lambda}$ and energy
density $\Lambda$. However, individual vacuum condensates known from
particle physics e.g. those which are responsible for the chiral and
gauge symmetries breaking in the Standard Model, contribute to the
vacuum energy of the Universe individually exceeding the observable
value of the $\Lambda$-term density $\Lambda_{\rm{exp}}=(3.0\pm
0.7)\times 10^{-35}\;\text{MeV}^4$ \cite{WMAP} by many orders of
magnitude in absolute value \cite{Martin:2012bt}. This situation,
which sometimes referred to as the ``Vacuum Catastrophe'' in the
literature, requires extra hypotheses about (partial or complete)
compensation of vacuum condensates of different types to the net
vacuum energy density of the Universe (see e.g.
Ref.~\cite{Polchinski:2006gy}). A dynamical mechanism for such gross
cancellations and corresponding major fine-tuning of vacuum
parameters is yet not known and is a subject of ongoing intensive
studies (for a review on this topic, see e.g.
Ref.~\cite{Copeland:2006wr} and references therein).

Within the general problem of vacuum condensates cancellation, the
QCD vacuum contribution has a special status. Various existing
cancellation mechanisms refer essentially to an unknown high-scale
physics beyond the Standard Model e.g. to Supersymmetry
\cite{Copeland:2006wr}. However, they cannot be applied for a
compensation of the specifically non-perturbative and low-energy QCD
contribution. In this paper, we focus primarily on elimination of
this most ``difficult'' part of the vacuum energy of the Universe.

In the framework of the popular instanton liquid models
\cite{instantons}, the topological (or instanton) modes of the QCD
vacuum (which sometimes referred to as the quark-gluon condensate)
are given essentially by the strong non-perturbative fluctuations of
the gluon and light sea quark fields which are induced in processes
of quantum tunneling of the gluon vacuum between topologically
different classical states. The topological instanton-type
contribution $\varepsilon_{vac(top)}$ to the energy density of the
QCD vacuum is one of its main characteristics \cite{Shifman:1978bx}
and can be written as follows (see also Ref.~\cite{Voloshin})
\begin{eqnarray}
\varepsilon_{vac(top)}&=&-\frac{9}{32}\langle0|:\frac{\alpha_s}{\pi}F^a_{\mu\nu}(x)
F_a^{\mu\nu}(x):|0\rangle + \frac14 \Big[\langle0|:m_u\bar
uu:|0\rangle+\langle0|:m_d\bar dd:|0\rangle \nonumber \\&+&
\langle0|:m_s\bar ss:|0\rangle\Big] \simeq -(5\pm 1)\times 10^{9}\;
\text{MeV}^4\,,
 \label{top}
\end{eqnarray}
which is composed of gluon and light sea $u,d,s$ quark
contributions. Clearly, other contributions of a different physical
nature should compensate the topological QCD contribution
(\ref{top}) to the vacuum energy of the Universe since its value by
far is not compatible with the cosmological observations and data on
the $\Lambda$-term value \cite{WMAP}. This issue triggers the search
for possible cancellation mechanisms, and one such mechanism will be
discussed further in this paper.

\section{Classical evolution of the cosmological Yang-Mills fields}

Consider one of the possible ways to eliminate the {\it microscopic}
QCD vacuum contribution (\ref{top}) to the vacuum energy density of
the Universe introducing the hypothesis about the existence of the
cosmological {\it macroscopic} Yang-Mills fields in early Universe.

Cosmological solutions for classical Yang-Mills fields have a long
history referring back to the late seventies, when there was an
active search for solutions to the Einstein-Yang-Mills field
equations \cite{YM-eqs}. Later, the role of non-Abelian gauge fields
in the early Universe evolution has been intensively studied in many
different aspects, in particular, in the context of the Dark Energy
\cite{YM-DE} and non-Abelian fields driven inflation without a
presence of a scalar field (``gauge-flation'') \cite{gauge-flation}.
Practically, there are no any physical arguments which could forbid
the existence of the homogeneous non-Abelian gauge field with
unbroken $SU(N)$ symmetry at cosmological scales with an isotropic
energy-momentum tensor \cite{Cembranos:2012ng}, possibly originating
from the inflationary stage of the Universe evolution
\cite{gauge-flation}.

Let us now assume that a chromodynamical (gluon) field with unbroken
color $SU(3)_c$ symmetry exists as a real physical object filling up
the early Universe, and the subject of our further discussion
concerns possible physical states of this field and their real-time
dynamics. For simplicity, we work in the flat Friedmann Universe
with conformal metric $g_{\mu\nu}=a^2(\eta)g_{\mu\nu(\rm M)}$, where
$g_{\mu\nu(\rm M)}$ is the Minkowski metric. The Einstein equations
with energy-momentum tensor of classical Yang-Mills fields are
\cite{YM-eqs}
\begin{eqnarray}
&&\frac{1}{\varkappa}\left(R_\mu^\nu-\frac12\delta_\mu^\nu
R\right)=\frac{1}{g_{\rm YM}^2}\frac{1}{\sqrt{-g}}\left(
-F_{\mu\lambda}^aF^{\nu\lambda}_a+ \frac14\delta_\mu^\nu
F_{\sigma\lambda}^aF^{\sigma\lambda}_a\right)\,,\quad \sqrt{-g}=a^4(\eta)\,, \nonumber\\
&&\left(\frac{\delta^{ab}}{\sqrt{-g}}\partial_\nu\sqrt{-g}-f^{abc}A_\nu^c\right)
\frac{F_b^{\mu\nu}}{\sqrt{-g}}=0\,,\quad F_{\mu\nu}^a=\partial_\mu
A_\nu^a-\partial_\nu A_\mu^a+f^{abc}A_\mu^bA_\nu^c\,.
\label{classicalYM}
\end{eqnarray}
This system is written in the most trivial form without taking into
account interactions of the macroscopic Yang-Mills field with the
physical vacuum (no vacuum polarisation effects are included here)
and other forms of matter (i.e. $\varepsilon=0$). Here and below,
raising and lowering Lorenz indices are done by the Minkowski metric
$g_{\mu\nu(\rm M)}$ as usual.

Since initial conditions in the early Universe are quite arbitrary,
it is meaningful to start with the study of spatially-homogeneous
and isotropic modes of the gluon field \cite{Cembranos:2012ng}. A
specific feature of such modes concerns their distinct topological
structure where the isotopic and spatial indices are mixed up. In
the case of Hamiltonian gauge $A_0^a=0$ and homogeneous and
isotropic 3-space we have the following simple structure of these
modes:
\begin{eqnarray}
\displaystyle A_i^a= \left\{ {\displaystyle \delta_i^aA(\eta),\quad
i,a=1,2,3
 \atop \displaystyle 0, \quad \quad  i=1,2,3;\; a>3\,,}  \right.
 \label{Afield}
\end{eqnarray}
with a single non-trivial time-dependent degree of freedom $A(\eta)$
to be studied in what follows. In this case, the classical
Yang-Mills equations (\ref{classicalYM}) read
\begin{eqnarray}
\frac{3}{\varkappa}\frac{a'^2}{a^4}&=&\frac{3}{2g_{\rm
YM}^2a^4}\left(A'^2+A^4\right)\,, \qquad A''+2A^3=0\,, \label{class}
\end{eqnarray}
and thus completely determine the conformal time evolution of the
spatially-homogeneous and isotropic Yang-Mills field. The second
equation in Eq.~(\ref{class}) can be exactly integrated, and its
general solution implicitly corresponds to non-linear oscillations,
i.e.
\begin{eqnarray}
A'^2+A^4=C^{4},\quad \int_{A_0}^A\frac{dA}{\sqrt{C^4-A^4}}=\eta,
\label{classol}
\end{eqnarray}
with $C,\,A_0$ being integration constants. Numerical solution of
Eq.~(\ref{class}) for the gluon field potential with initial
condition $A'(0)=0$ and an arbitrary amplitude $A_0=C$ to a good
accuracy can be approximated by
\begin{eqnarray}
A(\eta)\simeq A_0\cos\biggl(\frac65\,A_0\eta\biggr).
\label{classappr}
\end{eqnarray}
An essentially non-linear character of oscillations of the classical
YM field is thus emerged in explicit dependence of their amplitude
on frequency. According to Eqs.~(\ref{class}) and (\ref{classol}),
the spatially-homogeneous classical YM field in the isotropic
Universe behaves as an ultra-relativistic medium with energy density
$\varepsilon_{\rm YM}\sim 1/a^4$ and equation of state $p_{\rm
YM}=\varepsilon_{\rm YM}/3$ \cite{Cembranos:2012ng}.

\section{Role of the vacuum polarisation}

Can a classical Yang-Mills field be a component of the cosmological
medium in the radiation-dominated Universe? A simple analysis have
shown that the classical spatially-homogeneous Yang-Mills field
cannot exist in the early Universe since the classical Yang-Mills
equations (\ref{classicalYM}) are not form-invariant and unstable
with respect to radiative corrections. Such an instability emerges
due to the fact that there is no any threshold for vacuum
polarization of a massless non-linear gauge field, i.e. any
infinitesimally small external field is capable of reconstruction of
the classical Yang-Mills vacuum \cite{Savvidy}. Due to non-linearity
of initial operator Yang-Mills equations the vacuum polarization of
the massless quantum gluon field by its classical component leads to
a modification of classical equations. In practice, we deal with the
Savvidy equations for the Savvidy vacuum fluctuations \cite{Savvidy}
and look for their spatially-homogeneous modes. As was also stressed
in Ref.~\cite{Elizalde:2003ku}, similar quantum effects such as the
gluon condensation and the vacuum polarization effects can be
important for generation of an effective cosmological constant with
a negative equation of state in the system of coupled Born-Infeld
and gravitational fields in early Universe.

Let us analyze the Yang-Mills equations of motion incorporating the
vacuum polarisation effects. The Lagrangian of the gluon field
taking into account the vacuum polarisation in the one-loop
approximation has the following form \cite{Savvidy}:
\begin{eqnarray}\nonumber
L_{\rm YM}=
-\frac{11}{128\pi^2}\frac{F^a_{\mu\nu}F_a^{\mu\nu}}{\sqrt{-g}}\ln\biggl(\frac{J}{\Lambda_{\rm
QCD}^4}\biggr)\,,\qquad
J=\frac{1}{\xi^4}\frac{|F_{\alpha\beta}^aF^{\alpha\beta}_a|}{\sqrt{-g}}\,.
\label{Lagr}
\end{eqnarray}
Here, the numerical parameter $\xi$ is not fixed and reflects an
ambiguity in normalisation of the corresponding gauge/Lorentz
invariant $J$. Such a Lagrangian leads to a modified system of
equations for gravitational and Yang-Mills fields in the isotropic
Universe with vacuum polarisation effects incorporated, namely,
\begin{eqnarray}
&&\frac{1}{\varkappa}\left(R_\mu^\nu-\frac12\delta_\mu^\nu R\right)=
{T_\mu^\nu}^{,\,{\rm mat}} + \bar{\Lambda}\delta_\mu^\nu +
\frac{11}{32\pi^2}\frac{1}{\sqrt{-g}}\biggl[\biggl(-F_{\mu\lambda}^aF^{\nu\lambda}_a
\nonumber
\\
&& \qquad\qquad +\,\frac14\delta_\mu^\nu
F_{\sigma\lambda}^aF^{\sigma\lambda}_a\biggr)
\ln\frac{e|F_{\alpha\beta}^aF^{\alpha\beta}_a|}{\sqrt{-g}\,(\xi\Lambda_{\rm
QCD})^4}-\frac14 \delta_\mu^\nu \,
F_{\sigma\lambda}^aF^{\sigma\lambda}_a\biggr]\,,\label{modYM} \\
&&\left(\frac{\delta^{ab}}{\sqrt{-g}}\partial_\nu\sqrt{-g}-f^{abc}A_\nu^c\right)
\left(\frac{F_b^{\mu\nu}}{\sqrt{-g}}
\ln\frac{e|F_{\alpha\beta}^aF^{\alpha\beta}_a|}{\sqrt{-g}\,(\xi\Lambda_{\rm
QCD})^4}\right)=0\,, \nonumber
\end{eqnarray}
where $e\simeq 2.71$ is the base of the natural logarithm;
$\Lambda_{\rm QCD}$ is the QCD energy scale;
${T_{\mu}^{\nu}}^{,\,{\rm mat}}=(\varepsilon+p)u_\mu
u^\nu-\delta_{\mu}^{\nu} p$ is the energy-momentum tensor of all
components of the cosmological medium except for the macroscopic
Yang-Mills field; $\bar{\Lambda}=\Lambda_{\rm inst}+\Lambda_{\rm
cosm}+\dots$ is the total contribution to the vacuum energy density
which consists of the {\it non-perturbative} spatially-inhomogeneous
(topological) quantum fluctuations of the gluon and quark fields
(quark-gluon condensate) of an instanton nature (\ref{top}),
$\Lambda_{\rm inst}\equiv \varepsilon_{vac(top)} \simeq -(5\pm
1)\times 10^{9}\,{\rm MeV}^4$, an uncompensated contribution from
the observable cosmological $\Lambda$-term, $\Lambda_{\rm cosm},$
and dots represent all other {\it perturbative} vacua contributions.
The $\Lambda$-term value, $\Lambda_{\rm cosm},$ could have a
different nature, other than topological non-perturbative one in QCD
or perturbative ones in high-energy particle physics, so we
explicitly separated it from the rest. Form now on, we implicitly
assume that perturbative components of the net vacuum energy density
from all other microscopic vacuum condensates in particle physics
are compensated elsewhere at high energy scales and do not enter the
vacuum energy density of the Universe, so
$\bar{\Lambda}=\Lambda_{\rm inst}+\Lambda_{\rm cosm}$.

The system of equations (\ref{modYM}) is written in the most general
form including all forms of matter, as well as the uncompensated
quark-gluon condensate contribution $\Lambda_{\rm inst}$ and the
observable cosmological $\Lambda$-term $\Lambda_{\rm cosm}$. The
components of the energy-momentum tensor for the homogeneous and
isotropic modes specified in Eq.~(\ref{Afield}) have the following
generic form:
\begin{eqnarray}\nonumber
{T_0^0}^{,\,{\rm tot}}&=&{T_0^0}^{,\,{\rm mat}}+\bar{\Lambda}+
\frac{33}{64\pi^2}\frac{1}{a^4}\biggl[(A'^2+A^4)
\ln\frac{6e|A'^2-A^4|}{a^4(\xi\Lambda_{\rm
QCD})^4}+A'^2-A^4\biggr]\,,
\quad {T_{0}^{\beta}}^{,\,{\rm tot}}={T_{0}^{\beta}}^{,\,{\rm mat}}\,,\\
{T_{\alpha}^{\beta}}^{,\,{\rm tot}}&=&{T_{\alpha}^{\beta}}^{,\,{\rm
mat}}+ \bar{\Lambda}\delta_{\alpha}^{\beta}+
\frac{11}{32\pi^2}\frac{1}{a^4}\delta_{\alpha}^{\beta}
\biggl[-\frac12(A'^2+A^4)\ln\frac{6e|A'^2-A^4|} {a^4(\xi\Lambda_{\rm
QCD})^4}+\frac32(A'^2-A^4)\biggr] \label{TeI}
\end{eqnarray}
In flat and isotropic Universe, trace of the Einstein equations and
the equation of motion of the macroscopic gluon field read,
respectively,
\begin{eqnarray}
&&\frac{6}{\varkappa}\frac{a''}{a^3}=\varepsilon-3p + 4\bar{\Lambda}
+ {T_{\mu}^{\mu}}^{,\,{\rm YM}}\,,\quad {T_{\mu}^{\mu}}^{,\,{\rm
YM}}=\frac{33}{16\pi^2}
\frac{1}{a^4}\left(A'^2-A^4\right)\,,\label{sysYM-1} \\
&&\frac{\partial}{\partial\eta}\left(A'\ln\frac{6e|A'^2-A^4|}
{a^4(\xi\Lambda_{\rm QCD})^4}\right)
+2A^3\ln\frac{6e|A'^2-A^4|}{a^4(\xi\Lambda_{\rm QCD})^4}=0.
\label{sysYM-2}
\end{eqnarray}
It is straightforward to show that the $(0,0)$ Einstein equation
\begin{eqnarray}
\frac{3}{\varkappa}\frac{a'^2}{a^4}=\varepsilon + \bar{\Lambda} +
\frac{33}{64\pi^2}\frac{1}{a^4}
\left[\left(A'^2+A^4\right)\ln\frac{6e|A'^2-A^4|}{a^4(\xi\Lambda_{\rm
QCD})^4}+ A'^2-A^4\right] \label{int00}
\end{eqnarray}
is the exact first integral of the system of equations
(\ref{sysYM-1}) and (\ref{sysYM-2}), while the exact first integral
of second equation (\ref{sysYM-2}) is
\begin{eqnarray}
\frac{6e(A'^2-A^4)}{a^4(\xi\Lambda_{\rm QCD})^4}=1\,.
\label{sysYM-2-int}
\end{eqnarray}
The latter leads to a considerable simplification of the
energy-momentum tensor, namely,
\begin{eqnarray}
&&{T_0^0}^{,\,{\rm tot}}={T_0^0}^{,\,{\rm mat}}+\bar{\Lambda}+
\frac{33}{64\pi^2}\frac{(\xi\Lambda_{\rm QCD})^4}{6e}\,,\nonumber\\
&&{T_{\alpha}^{\beta}}^{,\,{\rm tot}}={T_{\alpha}^{\beta}}^{,\,{\rm
mat}}+\left(\bar{\Lambda}+ \frac{33}{64\pi^2}\frac{(\xi\Lambda_{\rm
QCD})^4}{6e}\right)\delta_{\alpha}^{\beta}\,. \label{TeI-sol}
\end{eqnarray}

Now we can observe an interesting possibility to eliminate the
microscopic {\it negative QCD contribution} to the vacuum energy,
$\Lambda_{\rm{inst}}$, by means of the {\it constant positive
contribution} from the spatially-homogeneous mode of macroscopic
gluon field. A small non-compensated remnant -- the observable
$\Lambda$-term, $\Lambda_{\rm cosm}$ -- can, in principle, have a
different nature which will be discussed in our forthcoming
publication. The corresponding condition for the $\Lambda_{\rm
inst}\simeq -265^4\,\mathrm{MeV}^4$ compensation
\begin{eqnarray}
\frac{33}{64\pi^2}\frac{(\xi\Lambda_{\rm QCD})^4}{6e}+\Lambda_{\rm
inst}=0,\quad \Lambda_{\rm QCD}\simeq 280\,\mathrm{MeV}\,,
\label{comp}
\end{eqnarray}
however, is not fully automatic; it is satisfied for a certain value
of the normalisation parameter only, $\xi\simeq 4$, which should be
constrained in a complete theory of the QCD vacuum. Therefore, in
principle, one succeeds to eliminate the huge negative contribution
from spatially-inhomogeneous non-perturbative quantum fluctuations
of the gluon field by means of a positive contribution from
fluctuations of spatially-homogeneous macroscopic gluon field. This
is achieved by fixing the remaining freedom in normalization of the
Yang-Mills invariant $J$ in the Lagrangian (\ref{Lagr}). As we will
see below, both mutually compensating contributions to the vacuum
energy density of the Universe have a common instanton nature.

\section{Cosmological evolution of finite-time instantons}

Together with the compensation condition (\ref{comp}) and the first
integrals (\ref{int00})  and (\ref{sysYM-2-int}), the resulting
system of equations (\ref{sysYM-1}) and (\ref{sysYM-2}) is
dramatically reduced to the following simple form:
\begin{eqnarray}
& &{}\frac{3}{\varkappa}\frac{a'^2}{a^4}=\varepsilon + \Lambda_{\rm
cosm},\label{sys-fin-1}
\\
& &{}A'^2-A^4=a^4\frac{(\xi\Lambda_{\rm QCD})^4}{6e}\,.
\label{sys-fin-2}
\end{eqnarray}
Notice that under the exact cancellation condition (\ref{comp}) the
cosmological (macroscopic) evolution of the Friedmann Universe given
by the scale factor $a=a(\eta)$ is now completely decoupled from the
microscopic evolution of the gluon field $A=A(\eta)$. The physical
time scale for the cosmological evolution is of the order of the
Universe age $t_{\rm cosm}\sim 1/H$ (in terms of the Hubble
parameter $H$), while the typical time scale for the Yang-Mills
field evolution is of the order of the hadronisation time $t_{\rm
hadr}\sim 1/\Lambda_{\rm QCD}$. So at present epoch the right hand
side of Eq.~(\ref{sys-fin-2}) can be taken to be constant in time to
a good accuracy, or more precisely, given by a classical solution of
the Friedmann equation (\ref{sys-fin-1}). In practice, this means
that the dynamical cancellation under the condition (\ref{comp})
and, hence, the decoupling of the QCD vacuum fluctuations from the
hot cosmological plasma have effectively happened at the end of the
hadronisation epoch in the early Universe evolution.

For convenience, let us rewrite the Yang-Mills equation
(\ref{sys-fin-2}) in terms of dimensionless time and gauge field as
follows
\begin{eqnarray}
\biggl(\frac{d\tilde{A}}{d\tilde{\eta}}\biggr)^2-\tilde{A}^4=1\,,\qquad
\tilde{A}=A\,\frac{(6e)^{1/4}}{\xi\Lambda_{\rm QCD}}\simeq
\frac{A}{2\Lambda_{\rm QCD}}\,,\qquad
\tilde{\eta}=\eta\,\frac{\xi\Lambda_{\rm QCD}}{(6e)^{1/4}}\simeq
2\Lambda_{\rm QCD}\,\eta\,, \label{modA}
\end{eqnarray}
for $\xi\simeq 4$. For simplicity, we have chosen the initial values
of the Yang-Mills field and the scale factor as follows:
\begin{eqnarray}
\tilde{A}(\tilde{\eta}_0=0)\equiv\tilde{A}_0=0,\qquad
a(\tilde{\eta}_0=0)\equiv a_0=1\,. \label{fixed-nu}
\end{eqnarray}
We can see now that, indeed, the time scale of the Yang-Mills field
fluctuations is essentially microscopic and corresponds to the
$\Lambda_{\rm QCD}$ energy scale. The equation (\ref{modA}) can then
be easily integrated, and its general solution can be written in the
following implicit form:
\begin{eqnarray}
\int_{\tilde{A}_0}^{\tilde{A}}\frac{d\tilde{A}}
{\sqrt{1+{\tilde{A}}^4}}=\tilde{\eta}\,, \label{ln-sol}
\end{eqnarray}
where $\tilde{A}_0$ is an integration constant. Notably, the
analytical solution (\ref{ln-sol}) taking into account the QCD
vacuum polarisation, in fact, differs from the classical Yang-Mills
solution (\ref{classol}) by sign in front of ${\tilde{A}}^4$ under
the square root only, having though a significant effect on its time
dependence. Moreover, since the solution (\ref{ln-sol}) was obtained
under the exact cancellation condition (\ref{comp}), it corresponds
to the minimal energy of the QCD system in the ground state of the
Universe and, hence, is physically preferable.

For the initial conditions given by Eq.~(\ref{fixed-nu})
(independently on a particular $\dot{a}_0$ value), the solution for
$\tilde{A}(\tilde{\eta})$ (\ref{ln-sol}) obeys the following
properties:
\begin{itemize}
\item Symmetry: $\tilde{A}(-\tilde{\eta})=-\tilde{A}(\tilde{\eta})$.
\item Periodicity: $\tilde{A}(\tilde{\eta}\pm T)=\tilde{A}(\tilde{\eta})$.
\item Continuous intervals and singularities:
$\tilde{A}(\tilde{\eta} \rightarrow \pm T/4)=\pm \infty\,.$
\end{itemize}

Most importantly, it is continuous only at a finite
microscopically-small time interval $T\sim 1/\Lambda_{\rm QCD}$ and
corresponds to spatially-homogeneous pulses of the gluon field
potential with a constant energy-density. An analogous effect in a
modified Maxwell-$F(R)$ gravity was observed in Ref.~\cite{Bamba}
where the large-scale magnetic fields are generated due to the
breaking of the conformal invariance of the electromagnetic field
through its non-minimal gravitational coupling.

We stress also that once the compensation condition (\ref{comp}) has
been satisfied at a particular moment in time, it holds true for any
later times, and the YM fields and large negative $\Lambda_{\rm
inst}$ disappear from the resulting equations and do not participate
in the Universe evolution any longer. We, therefore, arrive at
quasistationary regime when the Universe evolution is completely
determined by usual matter and uncompensated cosmological
$\Lambda$-term only.

What is the physical interpretation of the result (\ref{ln-sol})?
Obviously, such a solution with regular singularities does not have
a quasiclassical interpretation (except for a vicinity of the
midpoints of the continuous intervals where the gluon field
potential is small and slowly changing). In practice, we deal with a
sequence of quantum fluctuations of the YM field in time, or, in
fact, with {\it the finite-time instantons}. The creation and
annihilation of such finite-time instantons should have essentially
quantum nature like the QCD instantons. In order to regularize
singularities in the quasiclassical solution (\ref{ln-sol}) one
needs to turn into a complete quantum theory taking into account
quantum corrections due to e.g. fermion-antifermion pair creation
and annihilation processes in the early Universe.

\section{Conclusion}

From the group theory point of view, our proposal is analogical to
the standard instantons theory in QCD \cite{instantons}. In both
these cases one deals with the mapping of 3-space onto $SU(2)$
subgroup elements, so the analogy between the resulting instanton
solutions is rather close. Indeed, based on the quasiclassical
result (\ref{ln-sol}) only, one can naively conjecture that {\it the
cosmological evolution of the YM field emerges a sequence of quantum
tunneling transitions through the time barriers} represented by the
regular singularities in the solution (\ref{ln-sol}). We have
observed that the positive constant energy density of
spatially-homogeneous finite-time instantons in the early Universe
can be cancelled with the negative constant QCD contribution from
spatially-inhomogeneous gluon field fluctuations induced by a
similar quantum tunneling of the gluon field, but through spatial
(not time!) topological barriers between different classical vacua.
This exhibits a remarkable similarity and interplay between
instantons of different types in the early Universe evolution.
Besides the $\Lambda_{\rm QCD}$ scale parameter, a degree of such a
cancellation at the moment relies on one yet free dimensionless
normalisation constant $\xi$ which cannot be fixed by known
perturbative QCD theory, and thus should be fine-tuned for the exact
compensation to hold. This freedom must be eventually fixed by
non-perturbative QCD dynamics.

At the level of the Einstein equations, we have explicitly shown
that an arbitrary $SU(2)$ configuration of the cosmological
Yang-Mills field leads to a color neutral (``white'') contribution
to the energy-momentum tensor which is represented in the form of
Lorentz-invariant $\Lambda$-term. Thus, there is no any danger that
the cosmological Yang-Mills field leading to explicitly ``white''
observables upon averaging over all stochastic $SU(2)$
configurations would violate well-known symmetries of the QCD
theory, and quark fields cannot affect this picture.

From the point of view of quantum tunneling, the chain of quantum
fluctuations in a certain approximation can be considered as
non-linear oscillations. In order to regularize the infinite
end-points of the continuous time intervals within the
quasiclassical approach, one could therefore consider a continuous
smearing of the resulting fluctuations by means of a non-linear
continuous parameterization such as
\begin{eqnarray}
\tilde{A}_{appr}(\tilde{\eta})=\frac{1}{a\,\sin(\omega
\tilde{\eta})+b\,\dfrac{\cos^2(\omega \tilde{\eta})}{\sin(\omega
\tilde{\eta})}} \label{approx}
\end{eqnarray}
with adjustable parameters $a,\,b$ and $\omega$. This approximation
has been qualitatively compared to the exact quasiclassical solution
(\ref{ln-sol}) and approaches it in the limit of small $a\to0$,
while the classical non-linear solution (\ref{classol}) is reached
in the limit $a\to b$ (up to an arbitrary initial phase) with
$\omega$ being dependent on the initial amplitude. Thus, the
parameterization (\ref{approx}) represents a simple continuous
interpolation between the classical and quasiclassical solutions,
and can be used in practical calculations in the quasiclassical
limit of the theory.

To summarize, the space-time dynamics of colored quarks and gluons
has to be considered from the QCD confinement point of view. One
should emphasize that a formal singularity in the gluon field
potential (\ref{ln-sol}) simply means that a quark (and a gluon) in
the Universe cannot experience free motion in 3-space during the
time periods larger than the typical time scale of confinement,
$\sim \Lambda_{\rm QCD}^{-1}$. In this sense, our solution
(\ref{ln-sol}) reflects the ``time'' aspect of confinement. Besides
the extremely small $\Lambda$-term problem, the exact compensation
mechanism, proposed above, can be viewed as a manifestation of the
QCD confinement since there are practically no non-zeroth gluon
fields propagating at the length scales larger than the typical
hadron scale $\sim1$ fm, and they certainly disappear at
macroscopically large cosmological scales typical for modern
Universe. The perturbative higher order QCD corrections which affect
the QCD $\beta$-function are typically small and do not change the
qualitative picture described above. A deeper theoretical
investigation of these aspects is essential for both
non-perturbative QCD and Cosmology, and is planned for further
studies.

 \vspace{1cm}

{\bf Acknowledgments}

Stimulating discussions and correspondence with Sabir Ramazanov,
Johan Rathsman and Torbj\"orn Sj\"ostrand are gratefully
acknowledged. This work is supported in part by the Crafoord
Foundation (Grant No. 20120520).

\end{document}